
\input phyzzx

\overfullrule=0pt
\font\twelvebf=cmbx12
\nopagenumbers
\footline={\ifnum\pageno>1\hfil\folio\hfil\else\hfil\fi}
\line{\hfil January 1993}
\line{\hfil CU-TP-588}
\line{\hfil CERN-TH.6780/93}

\vskip 1in
\centerline{ \twelvebf   The Dual Formulation of  Cosmic
Strings and Vortices$^\dagger$}

\vskip 0.4in
\centerline{ \it Kimyeong Lee$^*$}
\vskip 0.25in
\centerline{ Physics Department, Columbia University }
\vskip 0.07in
\centerline{New York, New York 10027, USA}
\vskip 0.13in
\centerline{and}
\vskip 0.13in
\centerline{Theory Division, CERN}
\vskip 0.07in
\centerline{CH-1211 Geneva 23, Switzerland}
\vskip 0.6in
\centerline{\bf Abstract  }

We study four dimensional systems of global, axion and local strings.
By using the path integral formalism,  we derive the dual formulation
of these systems, where Goldstone bosons, axions and massive vector
bosons are described by antisymmetric tensor fields, and strings
appear as a source for these tensor fields. We show also how
magnetic monopoles attached to  local strings are  described in  the
dual formulation.   We conclude  with some remarks.

\vfill
\footnote{}{$\dagger$ This work is supported in part by U.S. Dept. of
Energy, the NSF Presidential Young Investigator Award, and the Alfred
P. Sloan Foundation.}
\footnote{}{$*$ E-Mail Address: klee@cuphyf.phys.columbia.edu}

\vfill\eject

\def\pr#1#2#3{Phys. Rev. {\bf D#1}, #2 (19#3)}
\def\prl#1#2#3{Phys. Rev. Lett. {\bf #1}, #2 (19#3)}

\def\np#1#2#3{Nucl. Phys. {\bf B#1}, #2 (19#3)}
\def\pl#1#2#3{Phys. Lett. {\bf #1B}, #2 (19#3)}

\REF\rLATT{ A. Savit, Rev. Mod. Phys. {\bf 52}, 453 (1980).}
\REF\Kalb{M. Kalb and P. Ramond, \pr{9}{2273}{74}.}
\REF\rLund{ F. Lund and T. Regge, \pr{14}{1524}{76}.}
\REF\rDual{  E. Witten
\pl{153}{243}{85}; A. Vilenskin and T. Vachaspati,
\pr{35}{1138}{87}.}
\REF\rDavis{ R.L. Davis and E.P.S. Shellard, \pl{214}{219}{88}}

\REF\rMagnus{R.L. Davis and E.P.S. Shellard, \prl{63}{2021}{89} ;
 R.L. Davis,  \pr{40}{4033}{89} and  R.L. Davis,
Mod. Phys. Lett. {\bf A5}, 955 (1990).}

\REF\rBen{B. Gradwohl, G. Kalberman, T. Piran and E. Berschtinger,
\np{338}{371}{90}; U. Ben-Ya'acov, \np{382}{597}{92} and \np{382}{616}{92}.}

\REF\rColeman{E.S. Fradkin and A.A. Tseytlin, Ann. of Phys. (New York)
{\bf 162}, 32 (1985); S. Coleman and K. Lee, \np{329}{387}{90}.}

\REF\rKim{Y. Kim and K. Lee, `Vortex dynamics in Self-Dual
Chern-Simons Higgs Systems,'  CU-TP-574, CERN-TH.6701/92 and hep-th
9211035  (1992).}

\REF\rAxion{C.G. Callan and J.A. Harvey, \np{250}{427}{85};
S.G. Naculich, \np{296}{837}{88}. }

\Ref\rNielsen{ H.B. Nielsen and P. Olsen, \np{61}{45}{73}.}

\REF\rSuga{ A. Sugamoto, \pr{19}{1820}{79}.}
\REF\rBowick{  T.J. Allen, M.J. Bowick and A. Lahir,
\pl{237}{47}{90} and   Mod. Phys. Lett. {\bf A 6}, 559 (1991).}

\chapter{Introduction}

We study a class of four dimensional field theories of a complex
scalar field and other fields with a  global or local abelian
symmetry.  These theories have global, axionic or  local  strings as
solutions in the symmetry broken phase.  We derive the dual
formulation of these theories by using the path integral. The dual
formulation  has  been extensively used to study the phase structure
of these theories,\refmark{1}  and the dynamics of cosmic strings and
superfluid vortices.\refmark{2\sim 7} However, the dual formulation of
strings has been derived usually by using the field
equations\refmark{4}  or the canonical transformations,\refmark{5}
making the whole situation somewhat unsatisfactory.

On the other hand, the dual formulation of the theory of a complex
scalar field has been derived in the path integral when there is no
vortex.\refmark{8} Recently, in path integral formalism   we have
derived the dual  formulation of three dimensional systems of
vortices to study the statistics of vortices in Chern-Simons Higgs
systems.\refmark{9}   Here we extend the idea of Ref. [9] to get
the dual formulation of four dimensional systems with cosmic strings.
While there have been a large literature about the dual formulation,
we feel our work is somewhat new and could be used to study quantum
feature of the string dynamics.

There are several advantages  of the dual formulation of cosmic
strings.  As the interaction between strings and other fields is
more explicit, one can understand the string evolution clearly.
A string or superfluid vortex  moving on a  background charge density
feels the so-called Magnus force. This Lorentz type force can be seen
directly in the dual formulation.\refmark{6,9} When the length  scale
of a string motion is lower than the string core size, one can obtain
an effective action which describes the string dynamics and its
interaction with low energy modes.

The plan of this paper is as follows. In Sec.~2  we study the theory
of a complex scalar field with a global abelian symmetry. In the dual
formulation, a global string appears as the source of an
antisymmetric tensor field, which represents the Goldstone boson. In
Sec.~3, we study the dual formulation of a theory
where global strings appear as axionic strings.  As shown in Ref.[10] the
electromagnetic charge is not conserved in  this theory without taking
into account  the chiral fermion zero modes on the string.
In Sec.~4, we study the Maxwell Higgs theory where there are
Nielsen-Olsen vortices, or local strings.\refmark{11} We derive the
dual formulation where the gauge  field is integrated out and is not
explicit. The antisymmetric field has a Higgslike coupling which leads
to its mass term. Magnetic monopoles attached to local strings are
described in the dual formulation. In Sec.5, we conclude with some remarks.

\chapter{Global Strings}

We consider the theory of a complex scalar field
 $\phi = fe^{i\theta} /\sqrt{2}$, whose lagrangian is given by
$$ {\cal L} = { 1\over 2}
(\partial_\mu f)^2 + {1\over 2} f^2 (\partial_\mu \theta )^2
 - U(f).  \eqno\eq $$
As the theory (1) is invariant under a global transformation,
$\theta \rightarrow \theta + {\rm constant}$, there are  conserved
current,
$$ j_\mu = f^2 \partial_\mu \theta, \eqno\eq $$
and  global charge,
$$ Q = \int d^3 r f^2 \partial_0 {\theta}.   \eqno\eq $$
The ground state of the energy functional for the systems we consider
is chosen to be a broken phase because of either the potential or a
background charge density. The low energy mode is then given by the
Goldstone bosons or sound wave.

To understand the quantum aspect, we use the  generating functional
$$ Z= <F| e^{-iHT}|I> = \int [fdf][d\theta]
\bar{\Psi}_F \exp \{ i \int d^3x {\cal L} \} \Psi_I,\eqno\eq $$
where $[f] \equiv \prod_x f(x)$ is the  Jacobian factor for the radial
coordinate of the scalar field.  The initial and final wave functions
$\Psi_{F,I}$ give the  necessary boundary conditions.

A given field configuration in the path integral could contain strings,
around each of which  the value of the $\theta$ field changes by $2\pi$ times
an integer. We can in principle split the  $\theta$  field into two parts,
$$ \theta(t,\vec{r})  = \bar{\theta}(t,\vec{r})
 + \eta(t,\vec{r}),\eqno\eq $$
where $\bar{\theta} $  describes a given configuration of vortices and
$\eta$ represents singlevalued fluctuations around the vortex
configuration. The energy density and the  complex scalar field $\phi$
should be singlevalued, or equivalently $\partial_\mu
\bar{\theta}$ and  $e^{i\bar{\theta}} $ should be so. Each string is
described by
parameterized positions, $\vec{q}_a(\sigma) $ or covariantly
$q_a^\mu(\tau,\sigma)$, where $\sigma^\alpha = (\tau, \sigma)$ is the
string world sheet coordinate. We choose $\sigma$ so that
$\bar{\theta}$ increases by $2\pi$ when one wraps the string in the
direction of increasing $\sigma$ with the right hand.

For a  straight string along $z$ axis, we know that $(\partial_x
\partial_y - \partial_y \partial_x )\bar{\theta} = 2\pi
\delta^2(\vec{\rho})$.  By covariantizing it, we get the antisymmetric
tensor vortex  current,
$$ \eqalign{ K^{\mu\nu}(x) &\ \equiv
\epsilon^{\mu\nu\rho\sigma}
\partial_\rho  \partial_\sigma \bar{\theta}  \cr
&\ = 2\pi  \sum_a    \int d\tau d\sigma (
 \dot{q}^\mu_a q'^\nu_a - \dot{q}^\nu_a q'^\mu_a )
\delta^4 (x^\rho  - q_a^\rho(\tau,\sigma)),\cr}
\eqno\eq $$
where the dot and prime indicate the  differentiation by
$\tau$ and $\sigma$, respectively. $K^{\mu\nu}$ is independent of
reparameterizations of $\sigma^\alpha$ up to sign and  satisfies the
conservation law, $\partial_\mu K^{\mu\nu} = 0$.
{}From Eq.(2.6),  we can get
$$ \eqalign{ \partial_i \bar{\theta}(t,\vec{r}) &\ =
\epsilon_{ijk}  \partial_j \int d^3 s\,\,\, { K^{0k}(t, \vec{s})
\over 4\pi |\vec{r} -\vec{s}| } , \cr
\partial_0 \bar{\theta}(t,\vec{r})  &\ = \epsilon_{ijk} \partial_i
\int d^3 s\,\,\,  { K^{jk}(t,\vec{s}) \over 8\pi |\vec{r}-\vec{s}| }  ,
\cr}
\eqno\eq
$$
by using a time-independent Green function. By integrating Eq.(2.7),
we get
$$ e^{i\bar{\theta}(t,\vec{r})} = \exp \biggl\{ i
\int^{\vec{r}}_{\vec{r}_0} d\vec{s} \cdot \vec{\nabla}
\bar{\theta} (t,\vec{s}) \biggr\},
\eqno\eq $$
where $\vec{\nabla} \bar{\theta} $ is given by Eq.(2.7) and
$\vec{r}_0$ is a reference point.
The exponent at the right hand side of Eq.(2.8) is multivalued but the
exponential is singlevalued.

The measure for the $\theta$  variable becomes
$$  [d\theta] = [d\bar{\theta}][d\eta]= [dq^\mu_a] [d\eta],\eqno\eq  $$
which means that we sum over singlevalued fluctuations around a given
configuration of strings  and then sum over all possible string
configurations. A typical string configuration would have
the creation,  annihilation, crossing,
exchange of strings. The Jacobian factor from $[d\bar{\theta}] $ to
$[dq_a^\mu]$ is complicated.
The periodicity of the $\theta$ variable affects  only  the
quantizations of both  global charge and  vorticity, due to the
gradient term $(\partial_\mu \theta)^2$ in the lagrangian.

In the generating functional $Z$, we can  linearize the $\theta$
kinetic term  by introducing an auxiliary vector field
$C^\mu$,
$$ \eqalign{ &\   \exp  i\int d^4x \bigl[ {1\over 2}f^2
(\partial_\mu \theta )^2 \bigr] \cr
&\ =
 \int [f^{-4} dC^\mu] \exp  i\int d^4x \biggl\{ -{1
\over 2 f^2} (C^\mu)^2 + C^\mu \partial_\mu \bar{\theta} + C^\mu
\partial_\mu \eta
 \biggr\}.
\cr} \eqno\eq $$
As $\eta$ is singlevalued, one can integrate over
  $\eta$ in the standard way, leading to
$$ \int [d\eta] \exp \biggr[ i\int d^4x C^\mu \partial_\mu \eta \biggl]
= \delta(\partial_\mu C^\mu).\eqno\eq $$

Now we introduce the dual antisymmetric tensor field $B_{\mu\nu}$ to satisfy
$$ \int [dC^\mu]  \delta(\partial_\mu C^\mu)...
 =\int [dC^\mu][dB_{\mu\nu}]    \delta(
  C^\mu - {1\over 2}
\epsilon^{\mu\nu\rho\sigma} \partial_\nu B_{\rho\sigma}  )...
\eqno\eq $$
and the dots denote the rest of the integrand.  There would be an infinite
gauge volume which can be taken care of later, but there is
no nontrivial
Jacobian factor as the change of variables  is linear.
By  using the fact that
$$  \epsilon^{\mu\nu\rho\sigma}
 (\partial_\mu \bar{\theta})\partial_\nu B_{\rho\sigma}
  = K^{\mu\nu} B_{\mu\nu} \eqno\eq$$
up to a singlevalued total derivative, we  can integrate
over $C^\mu$, resulting in the lagrangian,
$$
 {\cal L}_D = {1\over 2}(\partial_\mu f)^2  - U(f)
+{1\over 12  f^2} H_{\mu\nu\rho}^2 + {1\over 2}  B_{\mu\nu} K^{\mu\nu},
\eqno\eq $$
where  $H_{\mu\nu\rho } \equiv  \partial_\mu B_{\nu\rho} + \partial_\nu
B_{\rho\mu} + \partial_\rho B_{\mu\nu} $ is the field strength of
$B_{\mu\nu}$. Note that the kinetic term
for $B_{\mu\nu}$ has the standard normalization, e.g., $ (\partial_0
B_{12})^2 /2$. The dual lagrangian (2.14) is invariant under  a local
gauge symmetry, $B_{\mu\nu} \rightarrow B_{\mu\nu} + \partial_\mu
\Lambda_\nu - \partial_\nu \Lambda_\mu$.

The resulting path integral becomes
$$ Z = \int [f^{-3}df][dq^\mu_a]
[dB_{\mu\nu}] \bar{\Psi}_F  e^{i\int d^4x {\cal L}_D }
\Psi_I .\eqno\eq
$$
 One can  now introduce the gauge fixing terms for $B_{\mu\nu}$.
The initial and final states should be rewritten in dual variables.
The Goldstone boson is now  described by $B_{\mu\nu}$  and
strings appear as a  source for $B_{\mu\nu}$.

The mass  of vortices arises from the cloud of the $f,B_{\mu\nu}$ fields
surrounding them. The variation of $B_{0i}$ will lead to  a Gauss's law,
$$ - \partial_j({1\over f^2}H_{0ij}) +    K^{0i} = 0 ,
\eqno\eq $$
which would dictate the field cloud around vortices.
When the string of vorticity $n$ is lying on the $z$ axis, the $f$
field would vanish  like $f\sim \rho^n$ as one approaches the string
on the $xy$ plane. This can be seen directly from the $f$ equation in
the original formulation,  or from the $f$ and $B_{\mu\nu}$ equations
in the dual formulation. The classical relation between the original fields and
dual fields can be found  from the field equations from the
lagrangians at each step of the transformations.    They are related to
each other by
$$ f^2\partial^\mu \theta  =
{1\over 6} \epsilon^{\mu\nu\rho\sigma}
H_{\nu\rho\sigma} . \eqno\eq $$

Let us consider now the string dynamics briefly.
In the original formulation, the string motion is determined
by the field equations for $f, \theta$. In the dual formulation, it
seems  that there is be an equation of motion for the string
directly from the variation of $q_a^\mu(\tau,\sigma)$. This  is not
the case as we will see now. From the dual lagrangian (2.14), we get
for the variation of $\delta q_a^\rho$
$$ \delta {\cal L}_D = \sum_a {1\over 2} \int d\tau d\sigma
H_{\mu\nu\rho}( q_a)  (\dot{q}^\mu_a q'^\nu_a - \dot{q}^\nu_a q'^\mu_a)
\delta q^\rho_a . \eqno\eq $$
Since  $f^2 \partial_\mu \theta $ vanishes at the string,
Eq.(2.17) implies that the above variation vanishes. The
field equation obtained from the variation of the string position is
trivial. In the dual formulation, the string motion is
again governed by the field equations  of $f, B_{\mu\nu}$.
This is consistent with the picture that the kinetic energy
of vortices arises entirely from the field cloud around it and so the
dynamics of vortices should be determined by the field surrounding them.

This leads naturally to a question, whether there is an effective
action in terms of string position which describes the string
dynamics. We imagine the string dynamics whose energy scale is much
lower than that of the string core scale or the mass of $f$. As the
Goldstone boson is massless, we expects the effective action to
describe both strings and Goldstone bosons.  This assumption  cannot
be valid all the time as strings will annihilate each other. The
effective action is usually given as the Nambu action for  strings and the
action for the antisymmetric tensor field,\refmark{2,3,4}
$$ \eqalign{ S_{eff} = &\   \sum_a \mu_0 \int d^2 \sigma_a \sqrt{ -\gamma }
\cr
&\ + \int d^4 x  \{ {1\over 12 v^2} H_{\mu\nu\rho}^2 +
B_{\mu\nu} K^{\mu\nu} \} ,
\cr }
\eqno\eq  $$
where $v$ is the asymptotic value of $f$,  $\mu_0$ is the bare
string tension, and  $\gamma_a$ is the determinant of the induced metric
on the string,
$$ \gamma_{a\, \alpha\beta} = {\partial q_a^\mu \over \partial
\sigma^\alpha} { \partial q_{a\mu} \over \partial \sigma^\beta} ,
\eqno\eq $$
where $v$ is the vacuum expectation value of $f$.
The bare string mass per unit length $\mu_0$ comes from the string
core region. A cutoff of scale $m_f$ is necessary to make the string
self-energy finite. Note that the $f$ field does not approach
exponentially to its vacuum value at the spatial infinity. For a
straight string lying along
the $z$ axis, one can see  in the cylindrical coordinate
$(\rho,\varphi, z)$,
$$ f \rightarrow v - {1 \over m_f \rho^2} ,  \eqno\eq $$
as $\rho \rightarrow \infty $.  Since  there is no sharp
transition between  core and outside regions, the bare mass density and
the necessary cutoff are not clearly defined.  Hence, the effective
action (2.19) probably describes the string dynamics in order of
magnitude and needs  an improvement.

In addition the action should be modified when there is an background
charge density. The reason is that there is a sound wave of speed
$v_s$ rather than a Goldstone mode at the low energy. There would be
 an effective action for  sound wave and strings  when strings move
slower than the  sound speed. In addition, there
is a Magnus force on the string from the background charge.
While these are intersting questions, it will not be pursued here.

\chapter{Axionic Strings}

Let us now consider the case where the scalar field describes  an axion
field. To achieve this  we need a fermion whose mass term is chirally
generated by
our complex scalar field and a gauge boson which is coupled to the
fermion, so that the global abelian symmetry is broken by the anomaly.
When the fermion is very massive, integration over fermionic
field introduces an effective interaction between the gauge field and
the phase of complex scalar field.    For simplicity, we will
consider the case where the gauge symmetry is abelian.
( The detail
aspects  of anomalies, chiral zero modes and bosonization ideas in the
following discussion has appeared in Ref.[10].  We include the results
in Ref.[10] to understand the dual formulation  better.) The resulting
interaction between axion and two photons is  given by
$$ {\cal L}_{a\gamma\gamma}  = - { \theta \over 32  \pi^2 }
\epsilon^{\mu\nu\rho\sigma} F_{\mu\nu} F_{\rho\sigma} .  \eqno\eq
$$
This effective action is multivalued and not well defined at each
string as $\theta$ loses its meaning.

The equivalent singlevalued lagrangian to consider is
$$ {\cal L}_A  = {1\over 2} (\partial_\mu f)^2 + {1\over 2} f^2
(\partial_\mu \theta)^2 - U(f)
- {1\over 4e^2} F_{\mu\nu}^2 +    Z^\mu \partial_\mu \theta,
\eqno\eq $$
where the Chern-Simons current is
$$ Z^\mu = {1\over 8\pi^2} \epsilon^{\mu\nu\rho\sigma} A_\nu
\partial_\rho A_\sigma,  \eqno\eq $$
satisfying $\partial_\mu Z^\mu = {1 \over 32 \pi^2}
\epsilon^{\mu\nu\rho\sigma} F_{\mu\nu} F_{\rho\sigma} $.
The lagrangian (3.2)  is not gauge invariant as $Z^\mu$ is not.
The current from the lagrangian (3.2) is given by
$$ \eqalign { J^\mu_A &\ = {\delta {\cal L}_A \over \delta A_\mu} \cr
&\ = - {1\over 8\pi^2} \epsilon^{\mu\nu\rho\sigma} \partial_\nu \theta
F_{\rho\sigma} + {1\over 8\pi^2} K^{\mu\nu} A_\nu , \cr}
\eqno\eq $$
which is not conserved,
$$ \partial_\mu J^\mu_A = - {1\over 16 \pi^2} K^{\mu\nu} F_{\mu\nu} .
\eqno\eq
$$
When the string lies along the $z$ axis,
$K^{0z} = 2\pi \delta^2(\vec{\rho}) $ and $\partial_\varphi \theta =
1$ in the cylindrical coordinate. The current (2.4)  becomes
$$ \eqalign{ &\  J^{\rho}_A = - {F_{0z}  \over 4\pi^2 \rho}, \cr
&\ J^{z}_A = - {1\over 4\pi} A_0 \delta^2(\rho). \cr}
\eqno\eq $$
For a uniform electric field along $\hat{z}$, there is a radial
current moving to the string.

It turns out there is an additional degree of freedom to solve this
puzzle of gauge noninvariance. A chiral fermion zero mode lies along a
string, leading to an current, $J_\chi$,  such that the  total
current, $J_A^\mu + J_\chi^\mu$, is conserved. The effect of
this chiral fermion zero mode can be seen more directly by the
bosonization.\refmark{10}   With the string metric
$\gamma^{\alpha\beta}$ and the  antisymmetric tensor field
$\epsilon^{\alpha\beta} $, the lagrangian for the chiral boson
$\chi(\sigma^\alpha)$ on a string  is given  by
$$ \eqalign{ L_\chi = &\ {1\over 2} (\partial_\alpha \chi - {1\over
2\sqrt{\pi}}
A_\alpha)^2 - {1\over 2\sqrt{\pi}}\chi \epsilon^{\alpha\beta}
\partial_\alpha A_\beta \cr
&\ + (\gamma^{\alpha\beta} - \epsilon^{\alpha\beta} )
\lambda_\alpha (\partial_\beta \chi - {1\over 2\sqrt{\pi}}  A_\beta), \cr
} \eqno\eq $$
where the gauge field is evaluated at the string.
The chiral lagrangian is not  invariant under the gauge transformation,
$ \delta \chi = 2\sqrt{\pi} \Lambda$ and $\delta A_\alpha =
\partial_\alpha \Lambda$. (While it is trivial to introduce the world sheet
 metric on the string, let us use here the Cartesian coordinate.)

Let us put  the string on the $z$ axis,  and define the light cone
variables on the string, $x^{\pm } = {1\over \sqrt{2} } (t \pm z)$
with $ g_{+-} = g_{-+} = \epsilon_{+-} = -\epsilon^{+-} = 1$.
The field equations from Eq.(3.7) with $t= \tau, \,\, z = \sigma$,
imply  that $\lambda_\alpha$ can be chosen to be zero and that
$$ \partial_- \chi - {1\over 2\sqrt{\pi}} A_- = 0 , \eqno\eq $$
where $A_\pm = (A_t \pm A_z) /\sqrt{2}$.
The electric current due to the chiral bosons becomes
$$ \eqalign{ &\ J^+_\chi = {1\over 4\pi}  A_- , \cr
&\ J^-_\chi = - {1\over \sqrt{\pi}} \partial_+ \chi + {1\over 4\pi}
A_+ , \cr}
\eqno\eq $$
which is not conserved,
$$ \partial_\alpha J^\alpha_\chi = {1\over 4\pi}\epsilon^{\alpha\beta}
\partial_\alpha A_\beta .
\eqno\eq
$$

For the axionic string lying along the $z$ axis,
the sum of currents from Eqs. (3.4) and (3.9) becomes
$$ \eqalign{ &\  J^\rho = - {1\over 4\pi^2} {1\over \rho} F_{0z} ,\cr
&\ J^+ = 0 ,\cr
&\ J^- = \biggl(  - {1\over \sqrt{\pi} } \partial_+ \chi + {1\over
2\pi} A_+ \biggr) \delta^2(\rho), \cr}
\eqno\eq $$
which is conserved due to the field equation for $\chi$.
Note that the current is chiral on the string.
The combined action from  Eqs. (3.2) and (3.7) is then gauge
invariant, and so has the conserved electromagnetic current (3.11).

Let us consider the dual formulation of this lagrangian. We take the
similar steps as in Sec.2. We first split $\theta$ into $\bar{\theta}$
and $\eta$, and introduce $C^\mu$. Integrate  over $\eta$
to get
$$ \int [d\eta] \exp [i\int d^4x (C^\mu + Z^\mu)\partial_\mu \eta ]
= \delta( \partial_\mu (C^\mu + Z^\mu)) .  \eqno\eq $$
We again introduce $B_{\mu\nu}$ to solve the delta function,
$$ \delta (\partial_\mu (C^\mu + Z^\mu)) = \int [dB_{\mu\nu}]
\delta(C^\mu + Z^\mu  - {1\over 2} \epsilon^{\mu\nu\rho\sigma}
\partial_\nu B_{\rho\sigma} ) ... \eqno\eq $$

Integration over $C^\mu $ leads to the dual lagrangian,
$$ \eqalign{ {\cal L}_{AD} = &\ {1\over 2} (\partial_\mu f)^2 - U(f)  -
{1\over 4e^2} F_{\mu\nu}^2 \cr
&\ + {1\over 12
f^2} \tilde{H}_{\mu\nu\rho}^2 + {1\over 2} B_{\mu\nu} K^{\mu\nu}  ,
 \cr}
\eqno\eq $$
where $\tilde{H}_{\mu\nu\rho} \equiv  H_{\mu\nu\rho}  -
\epsilon_{\mu\nu\rho\sigma} Z^\sigma $. Since the Chern-Simons current
is not gauge invariant, the  field strength
$\tilde{H}_{\mu\nu\rho} $ is invariant under the electromagnetic gauge
transformation $\delta A_\mu = \partial_\mu \Lambda $ only if the
$B_{\mu\nu}$ field is also transformed as
$$\delta B_{\mu\nu} = {1\over 8\pi^2} \Lambda F_{\mu\nu} . \eqno\eq $$
The dual lagrangian (3.14)  is   not invariant under this gauge
transformation due to the last term.

In general, there will be a chiral boson $\chi_a$, lagrangian
multiplier $\lambda_{a\alpha}$ and lagrangian $ {\cal L}_{\chi_a}$
for each string $q_a$. When we add the dual action from Eq.~(3.14) and
these chiral actions, the total action is  gauge invariant. The
generating functional  has become
$$ Z = \int [f^{-3} df dB_{\mu\nu} dq_a^\mu][d\chi_a d\lambda_{a\alpha}]
\exp i\biggl\{ \int d^4x {\cal L}_{AD} + \sum_a \int d^2\sigma_a
{\cal L}_{\chi_{a}} \biggr\} .
\eqno\eq $$
The conserved electromagnetic current in the dual formulation is given
by
$$ \eqalign{ J^{\mu} = &\  {1\over 16\pi^2 f^2}
\tilde{H}^{\mu\nu\rho}F_{\nu\rho}  \cr
&\  + \sum_a \int d^2 \sigma {\partial q^\mu_a \over \partial
\sigma^\alpha} { (\gamma^{\alpha\beta} + \epsilon^{\alpha\beta})
\over 2}
( -{1\over \sqrt{\pi} } \partial_\beta \chi_a + {1\over 2\pi} A_\beta
(q_a)) . \cr}
\eqno\eq $$
The relation between the original field and dual fields outside strings
is given as
$$ f^2 \partial^\mu \theta = {1\over 6} \epsilon^{\mu\nu\rho\sigma}
 \tilde{H}_{\nu\rho\sigma} ,
\eqno\eq $$
and Gauss's law from the variation of  $B_{i0}$ becomes
$$ -\partial_j ({1\over f^2} \tilde{H}_{0ij}) + K^{0i} = 0 .
\eqno\eq $$

\chapter{Local Strings }

Let us now consider the dual formulation of the Maxwell Higgs systems.
Some aspects have been studied in Ref.[12,13]. The lagrangian is
$$ \eqalign{  {\cal L}_{M} = &\   -{1\over 4e^2}F_{\mu\nu}^2
 + {\lambda \over 8} \epsilon^{\mu\nu\rho\sigma}
F_{\mu\nu}  F_{\rho\sigma} + A_\mu J^\mu_{ext}  \cr
&\ + {1\over 2}(\partial_\mu f)^2
  + {1\over 2}f^2 (\partial_\mu \theta + A_\mu )^2
 - U(f) . \cr}
 \eqno\eq $$
We assume that there are magnetic monopoles. The gauge field can be
spilted into the monopole part  $\bar{A}_\mu$ and the single valued
rest ${\cal A}_\mu$. The magnetic monopoles are described by the
vector potential with a Dirac string or equivalently by the Wu-Yang
construction. When $f \neq 0 $,   $\partial_i \theta + \bar{A}_i$ is  gauge
invariant and well defined.
Suppose that there is no point around the monopole where $f=0$. Then
$\epsilon_{ijk} \partial_j (\partial_k \theta + \bar{A}_k) = B_i^{mon}
$ without a Dirac string, which is impossible. There  should be a
line attached to the monopole along which $f$ is zero and $\theta$
changes by $2\pi$ when one goes around it. This exactly the cosmicOA
string. In the  broken phase, the magnetic field from a magnetic monopole
is shielded by the Meissner effect, and is channeled by a local
string.

The Dirac string of $\partial_\mu \theta + A_\mu$
becomes a real local string attached to the magnetic monopole,
$$ \epsilon^{\mu\nu\rho\sigma} \partial_\rho (\partial_\sigma \theta +
A_\sigma) = K^{\mu\nu}_{mon} + {1\over 2} \epsilon^{\mu\nu\rho\sigma}
F_{\rho\sigma}^{mon} ,  \eqno\eq $$
where the monopole field satisfies
$$ {1\over 2} \epsilon^{\mu\nu\rho\sigma} \partial_\nu
F_{\rho\sigma}^{mon} =  m^\mu
\eqno\eq $$
with $m^\mu (x)  = \sum_b 2\pi   \int d\tau {ds^\mu_b \over d\tau }
\delta^4(x-s_b)  $ and the  string current $K^{\mu\nu}_{mon}$ would be
given by Eq.(2.6) with
the end point of the internal parameter $\sigma_0$ would describe the
monopole, $q_a^{\mu}(\tau, \sigma_0) = s_a^\mu$. This vortex current
$K^{\mu\nu}_{mon}$ is no longer conserved. By applying
$\partial_\nu$ to Eq.(4.2), we get
$\partial_\nu K^{\mu\nu}_{mon} = - m^\mu$.
Here we consider monopoles of the minimum charge allowed by the Dirac
quantization. For each magnetic monopole, there will be a attached
string.  The strings in the configuration space can be
closed, half-open with a monopole attached at one side, or open with
monopole and antimonopole attached at  both ends.

After some steps similar to those in  Sec.2, we get a lagrangian
$$ \eqalign{ {\cal L}_1  = &\  {1\over 2}(\partial_\mu f)^2  -U(f)  -
{1\over 4e^2 } F_{\mu\nu}^2
+ {\lambda \over 8}\epsilon^{\mu\nu\rho\sigma}F_{\mu\nu}
F_{\rho\sigma} + {\cal A}_\mu J^\mu_{ext}
\cr
&\ + {1\over 12 f^2 } H_{\mu\nu\rho}^2  + {1 \over 2}
 ( K^{\mu\nu} + {1\over 2}
\epsilon^{\mu\nu\rho\sigma}F_{\rho\sigma}^{mon})   B_{\mu\nu}  +
{1\over 2} \epsilon^{\mu\nu\rho\sigma}
\partial_\mu {\cal A}_\nu   B_{\rho\sigma}, \cr}
\eqno\eq $$
where the  string currents describes both open and closed strings and
we dropped the interaction term between monopoles and external charge.
In order to integrate over the singlevalued  gauge field ${\cal
A}_\mu$, we introduce  an antisymmetric tensor field $N_{\mu\nu}$
so that
$$ \eqalign{ \delta  ( &\ F_{\mu\nu} -
(\partial_\mu {\cal A}_\nu - \partial_\nu {\cal A}_\mu)-
F_{\mu\nu}^{mon}) \cr
&\ = \int [dN_{\mu\nu}] \exp
\{ i \int d^4 x {1\over 4}  \epsilon^{\mu\nu\rho\sigma}
N_{\mu\nu}[F_{\mu\nu}-(\partial_\mu A_\nu - \partial_\nu A_\mu) -
F_{\mu\nu}^{mon} ]\} \cr}
\eqno\eq $$
There is no nontrivial Jacobian factor.

Integration over ${\cal A}_\mu$ leads to
$$  {1\over 2} \epsilon^{\mu\nu\rho\sigma} \partial_\nu
(N_{\rho\sigma} - B_{\rho\sigma} )
-J^\mu_{ext} = 0,  \eqno\eq
$$
which is consistent only if $\partial_\mu J^\mu_{ext} = 0$. If $J^\mu$
has a dynamical origin so that it is not conserved identically, we
cannot integrate over $A_\mu$ without getting multivalued
$N_{\rho\sigma}$.  With ${1\over 2} \epsilon^{\mu\nu\rho\sigma}
\partial_\rho V_{\rho\sigma}^{ext} = J^\mu_{ext} $, we can express
$$ N_{\mu\nu}  = B_{\mu\nu} +  V_{\mu\nu} +  V_{\mu\nu}^{ext} ,
\eqno\eq $$
where $V_{\mu\nu} = \partial_\mu V_\nu - \partial_\nu V_\mu$.
A  point external electric charge  appears as a   magnetic monopole in
the  $V_{\mu\nu}^{ext}$ field.
When there is a unform background charge density,  we can choose the
gauge so that $V_{yz}^{ext} = 2xJ^0_{ext} $ or $
V_{\theta\varphi}^{ext}  = 2 rJ^0_{ext} $.
We change variables from $N_{\mu\nu}$ to $V_\mu$,  and then
$[dN_{\mu\nu}] = [dV_\mu]$.

Now we can integrate over $F_{\mu\nu}$ and get the dual lagrangian,
$$ \eqalign{ {\cal L}_{MD} = &\ {1\over 2}(\partial_\mu f)^2 - U(f) +
{1\over 12f^2 } H_{\mu\nu\rho}^2   - {e^2 \over 4(1+ \lambda^2 e^4)}
\tilde{B}_{\mu\nu}^2 \cr
&\ -{ \lambda e^4 \over 8(1+ \lambda^2 e^4)}
\epsilon^{\mu\nu\rho\sigma} \tilde{B}_{\mu\nu} \tilde{B}_{\rho\sigma}
+ {1\over 2}
B_{\mu\nu}( K^{\mu\nu} +
{1\over 2}\epsilon^{\mu\nu\rho\sigma} F_{\rho\sigma}^{mon} ) ,
 \cr}
\eqno\eq $$
where $\tilde{B}_{\mu\nu} \equiv B_{\mu\nu} + V_{\mu\nu}  +
V_{\mu\nu}^{ext}$. Note that the dual lagrangian is
invariant under the gauge transformation, $ \delta B_{\mu\nu}  =
\partial_\mu \Lambda_\nu - \partial_\nu \Lambda_\mu$ and $\delta V_\mu
= -\Lambda_\mu$ because of Eq. (4.2).  One can see the above
derivation is not affected even
when the  couplings $e, \lambda$ are  depending on spacetime, for example,
describing axionic domain walls. When there is no  Higgs field, we can just
drop $f, B_{\mu\nu}$ from Eq.(4.8)  and get a dual formulation  of the
Maxwell theory,
where  magnetic monopoles and electric charges have exchanged  their
role.
The generating functional is now
$$
Z = \int [f^{-3}df][dB_{\mu\nu}][dq^{\mu}_a][dV_\mu]
\bar{\Psi}_F \exp \biggl\{ i \int d^4 x {\cal L}_{MD}    \biggr\} \Psi_I.
\eqno\eq $$
The massive vector boson is now described  by $B_{\mu\nu}$.
We could include a kinetic term for magnetic monopoles.

Gauss's law from the variation of $B_{0i}$ is given by
$$ -\partial_j ({1\over f^2} H_{0ij}) + {e^2 \over 1 + \lambda^2 e^4}
\tilde{B}_{0i}  - {\lambda e^4 \over 2(1 + \lambda^2 e^4)}
\epsilon_{ijk} \tilde{B}_{jk}
+{1\over 2} \epsilon_{ijk} F_{jk}^{mon}   + K^{0i} = 0.
\eqno\eq
$$
Gauss's law from the variation of $V_0$ leads to
$$ {e^2 \over 1 + \lambda^2 e^4}  \partial_i \tilde{B}_{0i}
- {\lambda e^4 \over 2(1 + \lambda^2 e^4)} \epsilon_{ijk} \partial_i
\tilde{B}_{jk}  = 0 , \eqno\eq
$$
which is a consequence of Eq.(4.10). The constraint (4.10)
 should be satisfied by the field
configuration around strings and monopoles.
The classical relation between the original variables  and dual variables are
given by
$$ \eqalign{ &\ f^2(\partial^\mu \theta + A^\mu ) = {1\over 6}
\epsilon^{\mu\nu\rho\sigma} H_{\nu\rho\sigma} ,\cr
&\ F_{\mu\nu} + F_{\mu\nu}^{mon} = {e^2 \over 2(1 + \lambda^2 e^4)}
\epsilon_{\mu\nu\rho\sigma} \tilde{B}^{\rho\sigma} -
{\lambda e^4 \over 1 + \lambda^2e^4} \tilde{B}_{\mu\nu}. \cr }
\eqno\eq $$
When there is a nonzero  external background charge $J^0_{ext}$, the
lowest energy configuration would be such that  this external charge
is shielded completely  by the Higgs fields. In terms of the dual
fields, there will be nonzero $H_{123} = f^2(\dot{\theta} + A_0 +
A^{ext}_0) = - J^0_{ext}$.

There are two mass scales $m_f, m_A$ when there is no background
charge. When $m_f \gg m_A$, we expect an effective action for strings
and the massive vector bosons, which would be given trivially by
a simple  generalization of Eq. (2.19). When there are magnetic
monopoles, we have to think about the effective action for open
strings with  massive end points. When there are nonzero charge
density, we have to think about the Magnus force and lowenergy modes.
Again, there  are various questions discussed before.

\chapter{Conclusion}

We found the path integral derivation of the dual formulation for
various theories of abelian symmetry with strings. Goldstone bosons,
axions and massive
vector bosons are described by antisymmetric tensor fields of rank
two, while cosmic strings appear as the sources for these  tensor fields.
While the dual formulation has been extensively used before, our
derivation puts the dual formulation in a better footing and
we hope this leads  to a better understanding of the string or vortex
dynamics. In addition, we should note that it is trivial to generalize
our dual formulation in  curved spacetime and euclidean time. In
euclidean time, there is a subtlety related to the boundary condition
which fixes the conserved charge. (See  the second paper of Ref.[8])

However, there
are many open questions arising as we have more precise formulation to
start. What is an effective action which describes strings and low
energy modes of a given theory?
When there are nonzero unform background charge density, it is well
known that there is a Magnus force on strings and the particle
spectrum changes. In this case, what is the  low energy effective
action for strings,  low energy modes  and Magnus force in this case?
Clearly this question is related to the vortex dynamics in superfluid.
Finally, we have the dual formulation in the path integral and could
study the quantum dynamics of vortices.

\endpage

\refout
\end